\documentclass{aa}
\usepackage{graphicx}
\usepackage{txfonts}
\begin{document}
   \title{Star formation in the Vela Molecular Ridge}

   \subtitle{Large scale mapping of cloud D in the mm 
    continuum\thanks{Based on observations collected at the European 
    Southern Observatory, La Silla, Chile, program 71.C--0088}}

   \author{F. Massi \inst{1}
          \and
          M. De Luca\inst{2,3}
	  \and 
  	  D. Elia\inst{4}
	  \and 
          T. Giannini\inst{3}
          \and
          D. Lorenzetti\inst{3}
	  \and
      	  B. Nisini\inst{3}
          }

   \offprints{F. Massi}

   \institute{INAF - Osservatorio Astrofisico di Arcetri, Largo E.\ Fermi 5,
              I-50125 Firenze, Italy
              \email{fmassi@arcetri.astro.it}
         \and
             Dipartimento di Fisica, Universit\`{a} degli studi di Roma
             Tor Vergata, Via della Ricerca Scientifica 1, I-00133 Roma,
             Italy 
%             \thanks{The university of heaven temporarily does not
%                     accept e-mails}
	\and
	     INAF - Osservatorio Astronomico di Roma, Via Frascati 33, 
             I-00040 Monteporzio Catone, Roma, Italy
             \email{deluca,giannini,dloren,bruni@mporzio.astro.it}
	\and
	     Dipartimento di Fisica, Universit\`{a} del Salento,
             CP 193, I-73100 Lecce, Italy
              \email{eliad@le.infn.it}
             }

   \date{Received ; accepted }

   \abstract
{The Vela Molecular Ridge is one of the nearest 
intermediate-mass star forming regions, located within the galactic
plane and outside the solar circle. Cloud D, in particular, hosts 
a number of small embedded young clusters.}
% aims heading (mandatory)
{We present the results of a large-scale map in the dust continuum
at 1.2 mm of a $\sim 1\degr \times 1\degr$ area within cloud D.
The main aim of the observations was to obtain a complete census of
cluster-forming cores and isolated (both high- and
low-mass) young stellar objects in early evolutionary phases.}
% methods heading (mandatory)
{The bolometer array SIMBA at SEST was used to map the 
dust emission in the region
with a typical sensitivity of $\sim 20$ mJy/beam. This allows a
mass sensitivity of $\sim 0.2$ $M_{\sun}$.
%, assuming a dust temperature
%of 30 K, an opacity at 1.3 mm of $0.5$ g$^{-1}$ cm$^{2}$ and
%a distance of 700 pc. 
The resolution is $24\arcsec$, corresponding to
$\sim 0.08$ pc, roughly the radius of a typical young embedded cluster in
the region. The continuum map is also compared to a large scale map
of CO(1--0) integrated emission.}
% results heading (mandatory)
{Using the CLUMPFIND algorithm, a robust sample of 29 cores has been
obtained, spanning the size range $0.03 - 0.25$ pc and the mass
range $0.4 - 88$ $M_{\sun}$. The most massive cores are associated both
with red IRAS sources and with embedded young clusters, and coincide
with CO(1--0) integrated emission peaks. The cores are distributed according
to a mass spectrum $\sim M^{-\alpha}$ and a mass-versus-size relation
$\sim D^{x}$, with $\alpha \sim 1.45-1.9$ and $x \sim 1.1-1.7$. They
appear to originate in the fragmentation of gas filaments
seen in CO(1--0) emission and their formation is probably induced by
expanding shells of gas. The core mass spectrum is flatter than
the Initial Mass Function of the associated
clusters in the same mass range, suggesting
further fragmentation within the most massive cores. 
A threshold $A_{V} \sim 12$ mag seems to be required 
for the onset of star formation in the gas.}
% conclusions heading (optional), leave it empty if necessary
{}


   \maketitle
%
%________________________________________________________________

\section{Introduction}

CO(1--0) is the best tracer of molecular gas and the most efficient
transition to map large sky areas, 
but it fails to probe the densest molecular cores.
This is due both to its large optical depth and to the fact that CO freezes onto
dust grains at the highest densities. However, the availability of large arrays of
submillimetre/millimetre bolometers has recently provided a means to carry out
large surveys of entire star forming regions searching for protostellar and
starless cores with high sensitivity (see, e.\ g., Motte et al.\ 1998,
\cite{motte01}; Johnstone et al.\ \cite{john};
Kerton et al.\ \cite{kerton}; Mitchell et al.\ 2001;
Reid \& Wilson 2005; Mookerjea et al.\ 2004; Hatchell et al.\ 2005).

%Hence, there exists a need to identify nearby molecular clouds that are 
%active star forming regions in order 
All these studies testify to the need to obtain as completely as possible
a census of the youngest
stellar population down to the lowest masses, and to perform accurate studies of the
environment where young stars form.  This is particularly true for
 {\it massive} star forming regions, since they usually are farther away than
low-mass star forming regions, with obvious problems in terms of resolution. In this respect,
the Vela Molecular Ridge (VMR) has proved to be one of the most interesting targets. It is the
nearest giant molecular cloud complex hosting massive (although mostly intermediate-mass) star formation
after Orion, but, unlike Orion, it is located {\it within} the galactic plane, where most
of star formation in the Galaxy occurs. 
Its distance is probably within a factor 2 of 
that of Orion (700 pc vs.\ 450 pc), so the losses in terms of resolution
and sensitivity are not very large and are partly compensated by the bigger
observable area with the same instrumental field of view. 
Also, one can in this manner study many
different star forming sites with the same observational and environmental biases.

The VMR was first mapped in the $^{12}$CO(1--0) transition with low resolution
($\sim 30\arcmin$) by Murphy \& May (1991). They subdivided it into
four main regions, named A, B, C 
and D. Yamaguchi et al.\ (1999) presented new higher-resolution ($\sim 8\arcmin$) 
maps of the VMR in the $^{12}$CO(1--0) and $^{13}$CO(1--0) transitions, better evidencing 
the complex structure. The issue of distance
was discussed by Liseau et al.\ (1992), who found that clouds A, B and D are at $700 \pm 200$
pc. Liseau et al.\ (1992) and Lorenzetti et al.\ (1993) obtained the first data on
young stellar objects in the region measuring the spectral energy distribution (SED) for 
the associated IRAS sources. Massi et al.\ (1999, 2000, 2003) subsequently
found that the IRAS sources with $L_{\rm bol} > 10^{3}$ $L_{\sun}$ are associated with
young embedded clusters. 

Vela D appears to be the most efficient of the VMR clouds in producing
young embedded clusters. A look at Fig.~1 of Yamaguchi et al.\ (1999) shows
that the region consists of more than one cloud; that located at $\ell =264\degr,
b=0\degr$ contains at least 5 embedded clusters (see Fig.~1 of Massi et
al.\ 2003), all studied by Massi et al.\ (2000, 2003, 2006). With the aim of investigating
the mechanisms leading to this high efficiency, and the star formation history in
general, in 1999 we decided to make large scale maps 
of the cloud at mm wavelengths with the SEST. Observations in $^{12}$CO(1--0)
and $^{13}$CO(2--1) are presented in Elia et al.\ (\cite{elia}). 
In 2003, the same region was mapped in the
1.2-mm continuum emission with the bolometer array SIMBA. The SEST beam,
$24 \arcsec$, i.e.
$\sim 0.08$ pc at 700 pc, matches, or is smaller than, the typical spatial size
of cluster-forming cores and that
of the embedded star clusters in the region (see Massi et al.\ \cite{massi06}).  
A small subset of these data
has already been published in Giannini et al.\ (2005). This paper presents the
full results of the SIMBA observations. In Sect.~\ref{obsda}, observations and data
reduction are described. Sect.~\ref{resu} reports the most
significant results, that are then discussed in Sect.~\ref{discu}, 
and Sect.~\ref{conclu} summarizes our conclusions.

%__________________________________________________________________

\section{Observations and data reduction}
\label{obsda}

A $\sim 1\degr \times 1\degr$ region of Vela D was observed 
in the 1.2-mm continuum between May 23--26, 2003, using the
37-channel bolometer array SIMBA (Nyman et al.\ 2001) at the
Swedish-ESO Submillimetre Telescope (SEST) sited in La Silla, Chile.
At this wavelength, the beam HPBW is $24\arcsec$.
A total of 17 sky areas $18\arcmin \times 14 \arcmin$ 
(azimuth $\times$ elevation) in size were
mapped in the fast scanning mode, with a scanning speed of
$80 \arcsec$ s$^{-1}$. Each map was repeated 3 to 4 times, except
for 3 marginal regions. We always performed a skydip and pointing check
every $\sim 2$ hrs, while the focus was checked at the beginning of each
observing run and at sunset. The pointing was better than $\sim
5\arcsec$. The zenith atmospheric opacities remained in the ranges
$0.166-0.244$ (May 26) and $0.263-0.317$ (May 23). 

\begin{figure*}
   \centering
   \includegraphics[width=15cm]{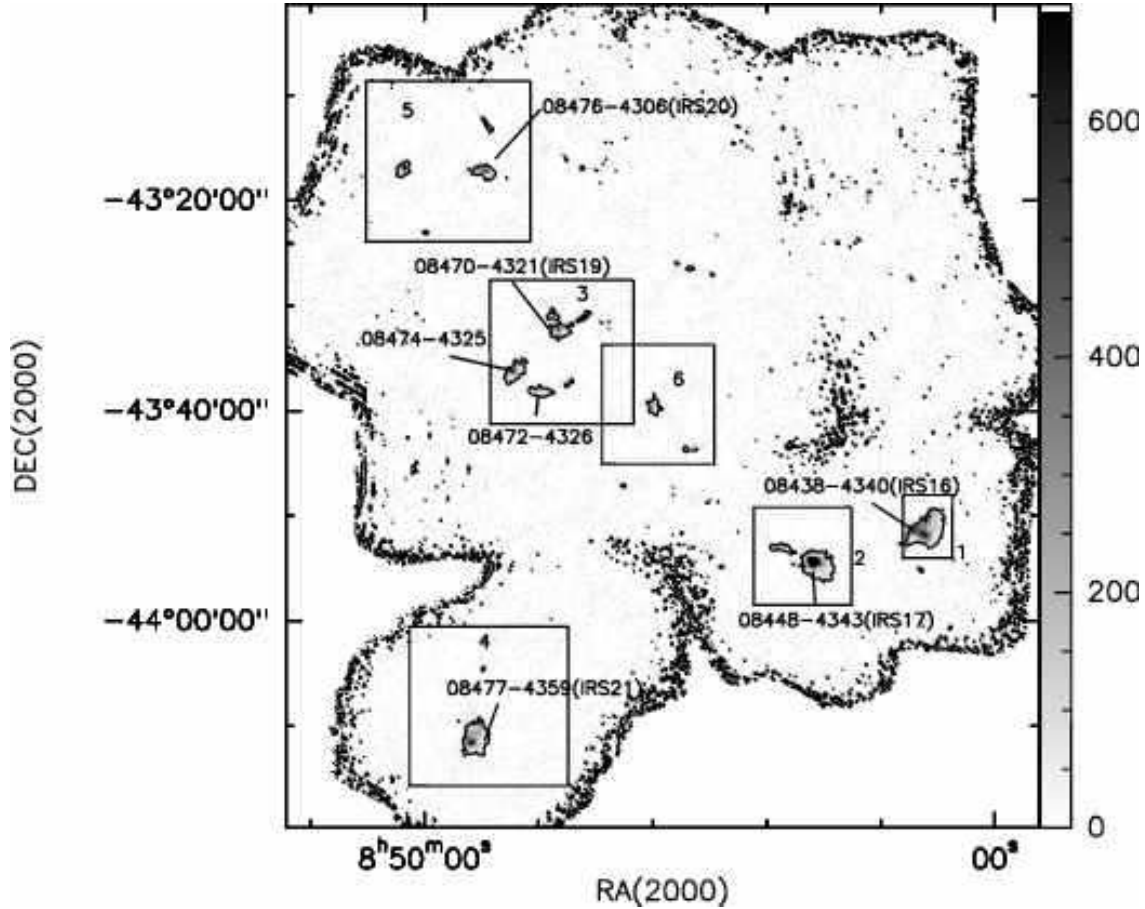}
      \caption{SIMBA 1.2-mm continuum map of cloud D (greyscale). 
      The scale (in mJy/beam) is indicated through the bar on the right and the
      contour at 60 mJy/beam ($\sim 3\sigma$) is also drawn.
      The red IRAS sources coincident with the most intense features
      are labelled (along with the designation adopted by Liseau et al.\ 1992).
      Six areas including the main mm sources are enclosed within boxes
      and numbered, and are shown zoomed-in in the following figures.  
         \label{fig:mm_map}
              }
   \end{figure*}

All data were reduced with MOPSI\footnote{MOPSI is a software
package for infrared, millimetre and radio data reduction developed
and regularly upgraded by R.\ Zylka.} according to the SIMBA
Observer's Handbook (2003). The steps are summarized in Chini et al.\
(2003). All the available maps for each of the 17 areas were first
coadded yielding 17 images that, once calibrated
(see below) were mosaiced together.
Since some of the images partially overlap, in those cases the
reduction steps were repeated while mosaicing in order to search for 
faint sources emerging from the areas with a longer effective integration
time.
The final large-scale map is shown in Fig.~\ref{fig:mm_map}. The r.m.s. is
in the range 14--40 mJy/beam, but is $\sim 20$ mJy/beam (or less) over most
of the map.

%
%   FIG. 1: MAPPONE (DA FARE)
%%%%%%%%%%%%%%%%%%%%%%%%%%%%%%%%%%%%%%%%%%%%
%%%%%%%%%%%%%%%%%%%%%%%%%%%%%%%%%%%%%%%%%%%%

Unfortunately, the only planet up during the observations was Jupiter,
which was mapped once per day.
Since it is resolved by the SEST beam, we decided then to calibrate our
1.2-mm data also using observations of Uranus carried out soon before and after
each of our observing runs. We derived the conversion factor from counts
to mJy/beam using different methods, 
e.g., by fitting a Gaussian to Uranus, by performing aperture photometry of
the calibrator using different apertures, by correcting the photometry of
Jupiter for source coupling and the error beam.
We obtained conversion factors within the range 55-75 mJy/beam counts$^{-1}$,
depending on the adopted method, but remarkably, each method provides 
a value that remains constant within a 20 \% (most often, within a 10 \%)
throughout the whole observing period (i. e., one week). 

Hence, we adopted the
canonical value of 65 mJy/beam counts$^{-1}$ for all runs but the last,
which always appears to approximate the measured values within 
$\sim 10$ \%. 
%
% INIZIO CORREZIONE 5
%
%Since there are hints of a slightly lower value
%strong hints of a slightly lower value
%during the last run (May 26), 
Since all our estimates of the conversion factor exhibit a small drop during
the last run (May 26),
%
% FINE CORREZIONE 5
%
in this case we adopted a value of 60 mJy/beam counts$^{-1}$.
Anyway, this choice does not affect the results, since only few faint sources
were detected during this run.
To check our data for consistency, we performed some aperture
photometry on a few sources using MOPSI. An example is the region
including IRS16 (here we will use the designation adopted by Liseau et al.\ 1992
for the IRAS sources common to their list) which 
was observed both on May 23 and on May 24.
The flux from the most intense source is found to be the same within $\sim 3$ \%,
confirming the excellent stability of the instrumental setting and atmospheric
conditions during the observations. In order to compare our data with the single pointing
measurements reported by Massi et al.\ (1999, 2003),
we roughly simulated single pointings
by selecting apertures with a diameter equal to the beam HPBW
and centred on the positions given by those authors.
For the eight common sources observed during our run,
%(IRS13, IRS17, IRS18, IRS19, IRS20, IRS21, IRS22, IRS31) 
we find flux densities within 36 \% of those measured by Massi et al.\ (1999, 2003);
in one case, the flux density value of Massi et al.\ (1999) is within  39 \% of ours.
However, from a check to the records of the observations listed in Massi et al.\ (1999) 
it seems that the single pointings are significantly affected by the selected throw. 
%However, note that this is only a rough comparison of measurements obtained in
%total different ways.

A much more significant test can be performed by considering different photometry
of sources observed in the same mode. Fa\'{u}ndez et al.\ (2004) mapped 
a sample of fields, which includes IRS17, with SIMBA in fast scanning mode.
They find a flux of 6.1 Jy for IRS17, while we find a total flux
of 8.6 Jy (see Giannini et al.\ 2005). However, we achieved an r.m.s. of $\sim 15$
mJy/beam, less than that of Fa\'{u}ndez et al.\ (2004), implying we might have
recovered more low-level emission. Fitting a Gaussian to the source, we obtained a size
of $44 \arcsec$ (same as Fa\'{u}ndez et al.\ 2004) and a flux of 5.5 Jy, within
$\sim 10$ \% of that measured by Fa\'{u}ndez et al.\ (2004).
Beltr\'{a}n et al.\ (\cite{beltran}) observed four of our sources in their survey
of massive protostar candidates with SIMBA, namely IRS16, IRS20 and IRS21 (plus
IRS18, observed in our run but located outside the target area, thus not discussed here).
They used CLUMPFIND to deconvolve the emission into clumps as we also did (see
Sect.~\ref{resu}). Since their maps are noisier and less sampled than ours,
the identified clumps are slightly different from the ones we detected. Hence, where
the number of clumps per source differs, we summed together the emission such as
to compare the same emitting areas. In two cases (IRS18 and IRS20), the two sets of fluxes
agree within $\sim 15$ \%. As for IRS16 and IRS21, the fluxes measured
by Beltr\'{a}n et al.\ (\cite{beltran}) are lower and within $\sim 50$ \% of ours.
We carried out aperture photometry with MOPSI on both sets of data, and found 
that the r.m.s. of the four maps by Beltr\'{a}n et al.\ (\cite{beltran}) is
twice as high as ours. 
%Their scanned areas are much smaller than 
%ours, also, and the S/N achieved for IRS16 and IRS21
%is relatively lower.
%
% INIZIO CORREZIONE 11
%
%, so that in these cases their integrated 
%fluxes and ours anyway agree within the errors.  
Their scanned areas are much smaller than ours, also, and
when using a small aperture around the most intense peaks, the fluxes from their frames
are found to be within $\sim 25$
\% of ours both for IRS16 and IRS21, confirming that probably we recover more low level
emission than Beltr\'{a}n et al.\ (\cite{beltran}). 
We conclude that our flux calibration is accurate in general to within $\sim 20$ \%.
%
% FINE CORREZIONE 11
%

\section{Results}
\label{resu}

%Stima della massa dai flussi a 1.2-mm, utilizzo di CLUMPFIND. I risultati
%sono mostrati in Tab.~\ref{table:cores}, che riporta identificazione dei clumps,
%posizione, flussi, massa e dimensioni. La tabella riporta i risultati
%definitivi ottenuti da Massimo e sar\'{a} referenziata da lui
%nell'articolo sulle cross-correlazioni.

\subsection{Dense cores}

Figure~\ref{fig:mm_map} shows a number of clumps throughout the region,
most of which are close to red IRAS sources. Among the young star clusters studied
by Massi et al.\ (\cite{massi00}, \cite{massi03}, \cite{massi06}), those that
are located within the boundaries of the observed area clearly coincide with
the brightest mm emission sources.

The map has been scanned, searching for dust condensations, using a
two-dimensional version of the tool CLUMPFIND, developed by Williams et al.\ 
(\cite{wi94}), capable of 
recognizing connected regions of emission with arbitrary shapes 
(hereafter {\em cores}), starting from the peaks of intensity and going 
down to the lowest contour levels depending on the r.m.s. of the data.

%-rms variabile ---> suddivisione in sottomappe
In order to reduce the influence of the variations in the r.m.s.
noise over
the map, we divided the observed region into 
ten smaller sub-regions. Then, we applied CLUMPFIND to each of
these areas using a local value of the r.m.s., as to properly set
the two input parameters: the lowest contour
level, $I_{\rm min}$, which acts as a detection threshold, and the intensity level
increment, $\Delta I$, used to separate resolved components located within the same
emission region. We in practise used $I_{\rm min} = 3$
r.m.s. and $\Delta I = 2$ r.m.s. in order to be conservative 
and to obtain a separation between substructures which seems reasonable
on a visual inspection.

The varying r.m.s., and the different input parameters chosen for each
sub-region, do affect the sensitivity over the map, but ultimately they  
only change the ``global'' completeness limit (estimated in 
Sect.~\ref{discu}). Above this completeness limit, we do not expect
our source statistics to be significantly biased.
% more than would result by, e.g., using a different finding algorithm. 

%       TAVOLA 1: CLUMPS TROVATI 
%_____________________________________________________________
%
%\begin{table*}
%\caption{Dust cores found by CLUMPFIND.}             
%\label{table:cores}      
%\centering          
%\begin{tabular}{c c c c c c c c c c c}     % 7 columns 
%\hline\hline       
%Core & \multicolumn{3}{c}{RA(J2000)} & \multicolumn{3}{c}{DEC(J2000)} & 
%Size & Peak & Integrated & Gas \\ 
% & & & & & & & & Flux & Flux & mass \\
% & & & & & & & (pc) & (mJy) & (mJy) & ($M_{\sun}$) \\
%\hline                    
%   1 & & & & & & & & & & \\  
%\hline                  
%\end{tabular}
%\end{table*}
\begin{table*}
\caption{Dust cores found by CLUMPFIND: the robust sample.
\label{table:cores}}
\centering
\begin{tabular}{c c c c c c c}     % 7 columns
\hline\hline
 Core & RA(J2000) & DEC(J2000) & Deconvolved & Peak & Integrated & Mass\\
 designation       &         &            & Size        & Flux &    Flux    & \\
       &          &          & (pc) & (mJy/beam)&    (Jy)    & ($M_{\odot}$) \\
\hline
 MMS1 & 8:45:33.4 & -43:50:20.1 & 0.19 & 501 & 3.17 & 32 \\
 MMS2 & 8:45:34.8 & -43:52:04.1 & 0.17 & 668 & 3.70 & 37 \\
 MMS3 & 8:45:40.1 & -43:51:32.3 & 0.19 & 544 & 3.59 & 36 \\
 MMS4 & 8:46:34.6 & -43:54:36.0 & 0.25 & 1702 & 8.79 & 88 \\
 MMS5 & 8:46:49.4 & -43:53:08.2 & 0.09 & 155 & 0.28 &  3 \\
 MMS6 & 8:46:52.3 & -43:52:59.9 & 0.24 & 121 & 0.20 &  2 \\
 MMS7 & 8:47:58.8 & -43:39:47.9 & 0.08 & 180 & 0.43 &  4 \\
 MMS8 & 8:48:39.4 & -43:31:23.9 & 0.02 & 77 & 0.08 & 0.8 \\
 MMS9 & 8:48:43.0 & -43:31:48.0 & 0.08 & 87 & 0.15 & 1.5 \\
 MMS10 & 8:48:43.0 & -43:37:08.0 & 0.03 & 83 & 0.13 & 1.3 \\
 MMS11 & 8:48:45.1 & -43:37:40.1 & 0.06 & 82 & 0.12 & 1.2 \\
 MMS12 & 8:48:49.0 & -43:32:28.0 & 0.13 & 617 & 1.77 & 18 \\
 MMS13 & 8:48:49.7 & -43:33:15.9 & 0.05 & 69 & 0.04 & 0.4 \\
 MMS14 & 8:48:51.1 & -43:31:08.1 & 0.08 & 117 & 0.12 & 1.8 \\
 MMS15 & 8:48:52.6 & -43:30:28.1 & 0.03 & 91 & 0.12 & 1.2 \\
 MMS16 & 8:48:53.3 & -43:31:00.1 & 0.05 & 163 & 0.23 & 2.3 \\
 MMS17 & 8:48:57.8 & -43:38:28.0 & 0.15 & 156 & 0.76 & 7.6 \\
 MMS18 & 8:49:03.6 & -43:38:12.1 & 0.09 & 105 & 0.28 & 2.8 \\
 MMS19 & 8:49:08.9 & -43:35:48.1 & 0.11 & 126 & 0.38 & 3.8 \\
 MMS20 & 8:49:11.8 & -43:35:24.0 & 0.09 & 126 & 0.19 & 1.9 \\
 MMS21 & 8:49:13.2 & -43:36:28.1 & 0.17 & 345 & 1.76 & 18 \\
 MMS22 & 8:49:27.4 & -43:17:08.2 & 0.09 & 567 & 1.32 & 13 \\
 MMS23 & 8:49:29.5 & -44:04:36.1 & 0.06 & 71 & 0.14 & 1.4 \\
 MMS24 & 8:49:31.0 & -43:17:08.2 & 0.06 & 421 & 0.76 & 7.6 \\
 MMS25 & 8:49:31.0 & -44:10:44.1 & 0.21 & 258 & 1.70 & 17 \\
 MMS26 & 8:49:33.8 & -44:10:59.9 & 0.20 & 298 & 1.55 & 15 \\
 MMS27 & 8:49:34.6 & -44:11:56.0 & 0.18 & 710 & 2.85 & 28 \\
 MMS28 & 8:50:09.4 & -43:16:27.8 & 0.06 & 118 & 0.20 & 2.0 \\
 MMS29 & 8:50:12.2 & -43:17:16.1 & 0.10 & 177 & 0.51 & 5.1 \\
\hline
\end{tabular}
\end{table*}
%
%
%
%_____________________________________________________________

%-criteri di ricerca dei clump (low, increment, npixmin ???)
%-output, stima masse e dimensioni
The CLUMPFIND output consists of a sample of about 50 cores for 
which we derived the
sizes by assuming a distance of $700$ pc. 
From the original output we selected a robust sample of 29 cores 
(listed in Table~\ref{table:cores}) which
fulfil the criterion of having a size (before beam deconvolution)
greater than, or equal to, the SIMBA HPBW. The sample does not include
those detections (although all are above 3 r.m.s.) that are revealed by CLUMPFIND, but 
``under-resolved'' because their size is less
than the SIMBA beam, {\em before} deconvolution. 
%Note that all sizes given in
%Table~\ref{table:cores} have been deconvolved.  
For completeness, we also list 
coordinates and peak intensity of these detections (estimates of sizes cannot be 
given) in Table~\ref{table:umms}. Most of them are small components ($\sol 1$ $M_{\sun}$) 
of large complex structures, but some (including isolated ones) appear in some
way linked to 
the star formation activity (e.g. coincidence with IRAS/MSX
point sources, CO peaks and/or evidence of H$_2$ emission), although we cannot exclude
they are artefacts of the finding algorithm. In particular, umms1 is a relatively 
high-mass core (but being located within a quite noisy area) and coincides with a
CO(1--0) peak at the border of the line map (hence only partially mapped). 
However, all these detections
deserve dedicated observations with higher sensitivity 
in order to confirm them, to clarify their nature and to give reasonable estimates of their
masses and sizes. A detailed analysis of these uncertain detections will be
presented in a forthcoming paper (De Luca et al.\ \cite{delu07}), 
while in the following we
will refer only to the sample of Table~\ref{table:cores}.
The adopted approach is bound to affect the statistic of the faintest sources,
dropping objects that would fall in the robust sample if the sensitivity were
higher. But, again, we do not expect major effects {\em above} the completeness
limit. 

The cores' 
masses have been determined accordingly from the expression of 
Fa\'{u}ndez et al.\ (\cite{faundez}), by assuming a gas to dust ratio of 100, 
a temperature 
of $30$ K  and a dust opacity $k_{1.3} = 0.5$ cm$^{2}$ g$^{-1}$ at $1.3$ mm.
We adopted the same $k_{1.3}$ as in Motte et al.\ (\cite{motte}) and Testi \& Sargent
(\cite{tes}); however, 
values as high as 1 cm$^{2}$ g$^{-1}$ are possible, typical of
very dense regions ($n_{\rm H} > 10^{7}$ cm$^{-3}$; Ossenkopf \& Henning \cite{osse}),
such as, e.g., circumstellar envelopes.
%This would result in masses that are a half of those computed with $k_{1.3} = 0.5$ 
%cm$^{2}$ g$^{-1}$. 

%
% TAVOLA 2
%
\begin{table*}
\caption{Possible dust cores found by CLUMPFIND, but with size
less than the SEST beam (before deconvolution). 
\label{table:umms}}
\centering
\begin{tabular}{c c c c c c c}     % 7 columns
\hline\hline
 Core & RA(J2000) & DEC(J2000) & Peak & Peak & Integrated & Mass \\
 Designation        &        &            & Flux & Flux &    Flux    &      \\
       &          &            & (mJy/beam) & ($\times$ local r.m.s.) & (Jy)  
& ($M_{\sun}$)\\
\hline
umms1 & 8:46:25.8 & -43:42:26.6 & 310 & 6 & 2.78 & 28  \\
umms2 & 8:46:37.2 & -43:18:34.9 & 86 & 4 & 0.05 & 0.5  \\
umms3 & 8:46:37.2 & -43:19:55.6 & 175 & 8 & 0.12 & 1.2  \\
umms4 & 8:46:49.0 & -43:20:27.1 & 111 & 5 & 0.09 & 0.9  \\
umms5 & 8:46:50.4 & -43:21:15.1 & 90 & 4 & 0.05 & 0.5  \\
umms6 & 8:46:56.6 & -43:53:07.1 & 97 & 6 & 0.16 & 1.6  \\
umms7 & 8:47:28.5 & -43:27:00.4 & 75 & 4 & 0.04 & 0.4  \\
umms8 & 8:47:37.3 & -43:43:40.1 & 71 & 4 & 0.03 & 0.3  \\
umms9 & 8:47:39.5 & -43:43:48.0 & 65 & 4 & 0.02 & 0.2  \\
umms10 & 8:47:41.0 & -43:26:28.3 & 84 & 5 & 0.12 & 1.2  \\
umms11 & 8:47:42.5 & -43:43:39.1 & 139 & 8 & 0.10 & 1.0  \\
umms12 & 8:47:46.8 & -43:25:55.4 & 79 & 5 & 0.03 & 0.3  \\
umms13 & 8:47:55.7 & -43:39:41.5 & 81 & 5 & 0.07 & 0.7  \\
umms14 & 8:47:57.9 & -43:38:59.6 & 110 & 7 & 0.17 & 1.7  \\
umms15 & 8:48:02.4 & -43:39:15.5 & 107 & 7 & 0.12 & 1.2  \\
umms16 & 8:48:15.7 & -43:47:07.8 & 92 & 5 & 0.06 & 0.6  \\
umms17 & 8:48:23.0 & -43:31:31.1 & 63 & 4 & 0.07 & 0.7  \\
umms18 & 8:48:26.7 & -43:31:39.7 & 66 & 4 & 0.04 & 0.4  \\
umms19 & 8:48:33.1 & -43:30:43.9 & 83 & 6 & 0.11 & 1.1  \\
umms20 & 8:48:35.5 & -43:30:59.8 & 83 & 6 & 0.07 & 0.7  \\
umms21 & 8:48:36.8 & -43:31:13.8 & 76 & 5 & 0.07 & 0.7  \\
umms22 & 8:48:36.6 & -43:16:51.2 & 84 & 4 & 0.05 & 0.5  \\
umms23 & 8:49:24.2 & -43:13:13.4 & 95 & 6 & 0.12 & 1.2 \\
umms24 & 8:49:27.1 & -43:12:33.5 & 86 & 6 & 0.07 & 0.7  \\
umms25 & 8:49:27.8 & -43:12:17.3 & 83 & 6 & 0.05 & 0.5  \\
umms26 & 8:49:59.0 & -43:22:55.6 & 91 & 6 & 0.09 & 0.9  \\
\hline
\end{tabular}
\end{table*}

By fitting 
%a grey-body emitter
(emissivity $\epsilon_{\lambda} \sim \lambda^{-1}$) the IRAS fluxes
at 60 and 100 $\mu$m, Liseau et al.\ (\cite{liseau}) find temperatures in
the range 30--43 K for IRS 17, 18, 19, 20 and 21. 
Fa\'{u}ndez et al.\ (\cite{faundez}) derive a typical temperature 
of 32 K for the cold dust component by a two-components fit to the 
SED's in their sample of southern high-mass star forming regions. 
Beltr\'{a}n et al.\ (\cite{beltran}) find a mean dust temperature
of 28 K by fitting the SED's in their sample of southern massive protostar
candidates longward of 60 $\mu$m. 
Hence, we adopted the canonical value of 30 K for all the
cores.  Nevertheless, some of the cores may be colder pre-stellar cores, and decreasing
the dust temperature to 15 K would increase the derived masses by a factor $2.5$. 
Actually, an external heating could drive the temperature of pre-stellar cores
towards higher values. Evidence of heating resulting 
from a strong external radiation field 
has been found for protostellar cores in Orion by J{\o}rgensen et al.\ (2006).
However, the presence of energetic sources in the mapped region of Vela-D
is still controversial (Lorenzetti et al.\ 1993, Elia et al.\ \cite{elia}). 

%-criteri di selezione e robust sample
The deconvolved sizes and the masses  of the cores 
listed in Table~\ref{table:cores} range 
from 0.03 to 0.25 pc and 0.4 to 88 $M_{\sun}$ respectively, while their 
spatial distribution exhibits a high degree of clustering.
The derived sizes span the interval from the observed diameter of pre-stellar cores in
cluster-forming regions to that of cluster-forming cores (or isolated
prestellar cores). The achieved typical sensitivity of $20$ mJy/beam translates
into a point source mass sensitivity (at a 1 $\sigma$ level) of 
$\sim 0.2$ $M_{\sun}$, using the above adopted temperature, opacity and distance. 
However, see the next section for an assessment of
the mass sensitivity to extended sources.

%-clump scartati (necessita maggiore risol.)
\subsection{Contamination sources of dust emission}

%While the effect of non-uniform dust temperature and opacity throughout the
%sample will be discussed in the next section, the other possible cause of
%error in the derived masses is contamination by line emission and/or
%free-free and synchrotron radiation. 
While the effect of non-uniform dust temperature and opacity throughout the
sample will be discussed in the next section, the other possible causes of
error in the derived masses are contamination by: i) line emission, ii) 
optically-thin free-free emission, iii) optically-thick free-free emission
and iv) synchrotron radiation. In the following, we discuss each of the
contamination sources. 

The line that can mostly affect 
the detected signal is $^{12}$CO(2--1). We estimated its contribution
using the relation given by Braine et al.\ (\cite{braine}), adapted
to our instrumental parameters, and our
observations of CO(1--0). Since CO(1--0) and CO(2--1) are optically thick,
assuming the same excitation temperature yields a CO(2--1) to CO(1--0)
ratio $\leq 1$. Due to the larger beam at the frequency of CO(1--0) with 
respect to that at 1.2 mm, by assuming the same main beam temperature for
CO(2--1) and CO(1--0) we still might slightly underestimate the CO(2--1)
main beam temperature. However, we checked that the line contribution
to the integrated flux density is always less than few percent for the
cores with $S_{\nu} \sog 10^{3}$ mJy. Conversely, for the cores with 
$S_{\nu} \sol 10^{3}$ mJy, the estimated line contribution is $\sim 10-30$ \%
(55 \% in one case). This is an overestimate, since the integrated emission
of CO(1--0) is $< 60$ K km s$^{-1}$ towards most of these sources and, as can be seen
in Fig~\ref{fig:co_mm_map}, at least the integrated emission up to 
25 K km s$^{-1}$ arises from a distributed structure much larger than the
cores' size that has been effectively filtered out by the observing
mode of SIMBA 
(see also Fig.~\ref{fig:irs17} to Fig.~\ref{fig:center}).

Hence, line emission is very likely not to contribute more than 10-20
\% of the integrated flux density from the faintest 1.2mm
sources. 
These results agree with what extensively discussed by Johnstone et al.\ (2003);
they find that at 850 $\mu$m, only occasionally the continuum becomes 
contaminated by line emission, including CO(3--2),
in strong sources, although this may be a problem
for low continuum flux sources.

Free-free emission from HII regions can also affect the measured flux densities
at 1.2 mm. As a first step, we checked that none of the cores' positions
coincide with radio sources from the Parkes-MIT-NRAO survey at 4.85 GHz
(Griffith \& Wright \cite{gew}). Hence, their radio fluxes at 4.85 GHz are lower
than the quoted limit of 48 mJy. For optically thin free-free sources 
this translates into 32 mJy at 1.2 mm. Only two of the cores listed in
Table~\ref{table:cores} exhibit an integrated flux density $< 100$ mJy,
whereas for all those with integrated flux density $\sim 10^{2}$ mJy the value
extrapolated from the radio upper limit amount to $< 25$ \% of
that measured at 1.2 mm.  As shown by
Massi et al.\ (\cite{massi03}), at a distance of 700 pc this upper limit
imply no HII regions excited by stars earlier than B1. Although we
assumed optically thin emission, this agrees with the limit on the earliest
star spectral type set by the observed bolometric luminosities (B1-B2). 

The Parkes-MIT-NRAO catalogue lists extended radio emission that, generally,
lies a few arcmin from the cores. However, there are some exceptions.
Towards IRS 16: MMS1, MMS2 and MMS3 are at a distance 
of $\sim 1\arcmin$ from the centre
of a radio source (quoted radius $1.1 \arcmin$) at
RA(J2000)=08:45:$36.3$, DEC(J2000)=$-43$:51:01. This is the HII
region $263.619$--$0.533$ (Caswell \& Haynes \cite{ceh}); the extrapolated
flux density (assuming optically thin emission) at 1.2 mm is $1.6$ Jy,
15 \% of the total integrated flux density of MMS1, MMS2 and MMS3. Assuming
the radio emission is uniform over a circle of radius $1.1 \arcmin$,
the expected flux density at 1.2 mm is 52 mJy/beam(SEST), i.\ e., at
a $2-3 \sigma$ level. Clearly, as can be seen in Fig.~\ref{fig:irs16},
the three cores are located at the boundary of the HII region (whose centre
almost coincides with the IRAS location and the centre of the radio emission),
confirming that they are mostly due to dust thermal radiation. 
Towards IRS19: MMS19, MMS20 and MMS21 lie $\sim 2 \arcmin$
from the centre of a radio source (PMNJ0849--4335; semi-major by semi-minor 
axes $3\farcm 8
\times 1\farcm 4$) at RA(J2000)=08:49:$22.6$, DEC(J2000)=$-43$:35:56.
The flux density extrapolated at 1.2 mm (assuming optically thin free-free radiation)
is 294 mJy, with a mean flux density of $2$ mJy/beam(SEST), well below our r.m.s.
This explains why no emission is detected at 1.2 mm from this radio source.
Anyway, the extrapolated total flux density is $\sim 13 \%$ of that measured
at 1.2 mm for the three cores. 
Finally, towards IRS21: MMS25, MMS26 and MMS27 are located $2-3\arcmin$
from the centre of a radio emission (PMNJ0849--4413;
semi-major by semi-minor axes $2\farcm 6
\times 1\farcm 5$) at RA(J2000)=08:49:$31.2$, DEC(J2000)=$-44$:13:47. 
The flux density extrapolated at 1.2 mm (assuming optically thin free-free radiation)
is 94 mJy, with a mean flux density of $1$ mJy/beam(SEST), again
well below our r.m.s.. In fact, no mm emission is detected from this
radio source, as well. The total extrapolated flux density is
only $< 2$ \% of that measured at 1.2 mm for the three cores. 

Clearly, on the one hand 
the contamination to the cores' flux densities from optically
thin free-free emission is negligible. Synchrotron radiation has an even
steeper spectral index than free-free ($S_{\nu} \sim \nu^{-0.6}$ vs.\
$S_{\nu} \sim \nu^{-0.1}$), so its contribution at 1.2 mm is expected to be
much lower. On the other hand, we cannot exclude some contamination from
optically thick free-free emission ($S_{\nu} \sim \nu^{2}$ up to a
turnover frequency depending on the emission measure) arising from 
ultracompact (or hypercompact) HII regions towards some of the cores.
Indeed, as discussed by Massi et al.\ (\cite{massi03}), some of the
IRAS sources towards the cores have the colours of UCHII regions,
but the fraction of fields exhibiting other signposts of UCHII regions
appears to be very low. 
 
\subsection{Star forming regions}

%                FIGURA 2: IRS 16 
%-------------------------------------------------------------
   \begin{figure}
   \centering
   \includegraphics[width=9cm]{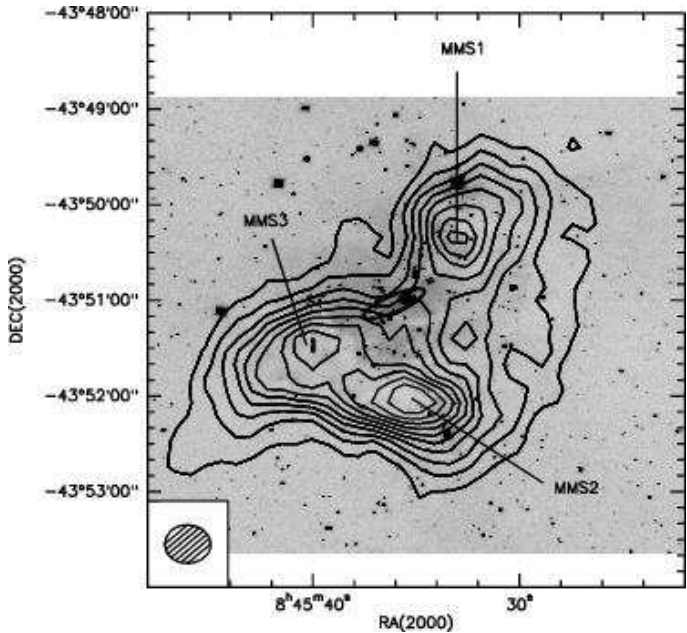}
      \caption{NTT/SofI image of IRS16 (at $K_{s}$), taken
      on December 2005, overlaid with a contour map
      of the 1.2-mm continuum emission (area 1 in Fig.~\ref{fig:mm_map}). 
      Contours are in steps
      of 60 mJy/beam ($\sim 3\sigma$) from 60 to 600 mJy/beam.
      The cores are labelled and the IRAS uncertainty ellipse is
      also drawn. The SEST beam is displayed in the lower left box. 
%
% CORREZIONE 6
%
	The nebulosity surrounding the location of the IRAS source
	is due to the HII region $263.619$--$0.533$.
%
% FINE CORREZIONE 6
% 
         \label{fig:irs16}}
   \end{figure}
%
%_____________________________________________________________

\begin{figure}
   \centering
   \includegraphics[width=9cm]{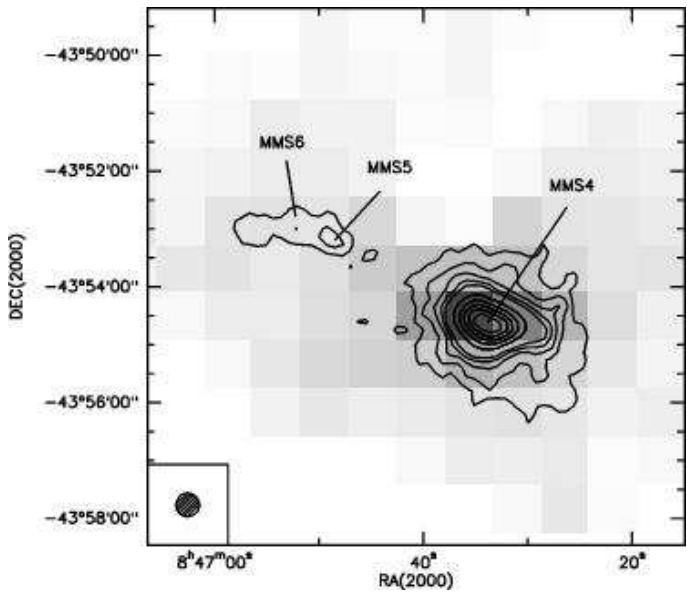}
      \caption{Contour map
      of the 1.2-mm continuum emission from the region including IRS17
      (area 2 in Fig.~\protect\ref{fig:mm_map}\protect). 
      Contours are in steps
      of 60 mJy/beam ($\sim 3\sigma$) from 60 to 300 mJy/beam 
      and in steps of 240 mJy/beam from 540 to 1500 mJy/beam.
      The cores are labelled and the SEST beam is
      drawn in the lower left box. 
      Underlying, the $^{12}$CO(1--0) emission integrated
      from 0 to 20 km s$^{-1}$ (greyscale, from 30 to 250
      K km s$^{-1}$). 
         \label{fig:irs17}}
   \end{figure}

Clues about the star formation history of the cloud can be derived
from the filamentary morphology of the gas (see Elia et al.\ \cite{elia}).
As clearly shown in Fig.~\ref{fig:co_mm_map}, all mm cores are aligned
along filaments or arcs of molecular gas. IRS16 and IRS17 lie inside
one of these arcs, in the south-western part of the map. IRS16 (see
Fig.~\ref{fig:irs16}) consists of 3 massive cores located around an HII
region (roughly $0.4$ pc in diameter), suggesting that the expansion of 
the HII region may have 
triggered star formation there. On the other hand, IRS17 
(see Fig.~\ref{fig:irs17}) is part of 
a filament of mm sources; it contains
the most massive core (MMS4) of the whole region, which is slightly elongated 
in the direction of the larger-scale surrounding CO arc. Two more
cores (MMS5 and MMS6), east of it and again following
the molecular gas orientation, suggest fragmentation of a larger filament. 
Giannini et al.\ (\cite{giannini}) showed that MMS4 can be further 
decomposed into two cores (mmA of $110$ $M_{\sun}$, and mmB of 11
$M_{\sun}$) embedded in more diffuse gas (58 $M_{\sun}$). They adopted a
smaller dust temperature ($23$ K), finding higher masses than ours. 

%_____________________________________________________________

%                FIGURA 3: IRS 17 
%-------------------------------------------------------------
%
%_____________________________________________________________

%_____________________________________________________________
%                FIGURA 4: IRS 19 
%-------------------------------------------------------------
   \begin{figure}
   \centering
   \includegraphics[width=9cm]{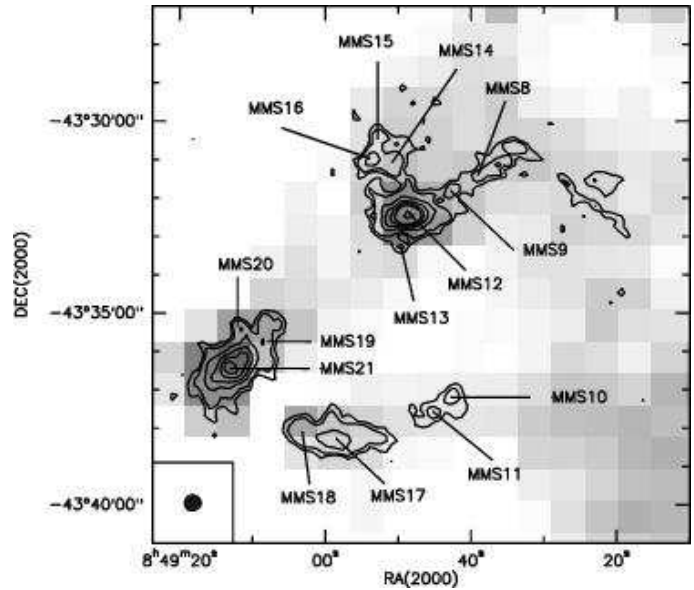}
      \caption{Contour map
      of the 1.2-mm continuum emission from the region including IRS19
      (area 3 in Fig.~\protect\ref{fig:mm_map}\protect).
      Contours are: 40 mJy/beam ($\sim 2\sigma$) the lowest one, then in steps
      of 60 mJy/beam ($\sim 3\sigma$) from 60 to 300 mJy/beam
      and 540 mJy/beam the highest one.
      The cores are labelled and the SEST beam is
      drawn in the lower left box.
      Underlying, the $^{12}$CO(1--0) emission integrated
      from 0 to 20 km s$^{-1}$ (greyscale, from 30 to 200
      K km s$^{-1}$). 
         \label{fig:irs19}}
   \end{figure}
%
%_____________________________________________________________

IRS19 (see Fig.~\ref{fig:irs19}) is part of a relatively complex
structure composed of filaments arranged in a roughly elliptical
pattern with a major axis of $\sim 2$ pc. 
The most massive cores in the pattern (MMS12, MMS17 and MMS21)
are all associated with three IRAS sources whose SED is typical 
of Class I sources (IRAS08470--4321/IRS19,
IRAS08472--4326, and IRAS08474--4325, respectively). This structure is 
reflected in the CO(1--0) map (Fig.~\ref{fig:co_mm_map}) also, 
suggesting a dynamical origin and a subsequent fragmentation.
 
%_____________________________________________________________
%                FIGURA 5: IRS 21 
%-------------------------------------------------------------
   \begin{figure}
   \centering
   \includegraphics[width=9cm]{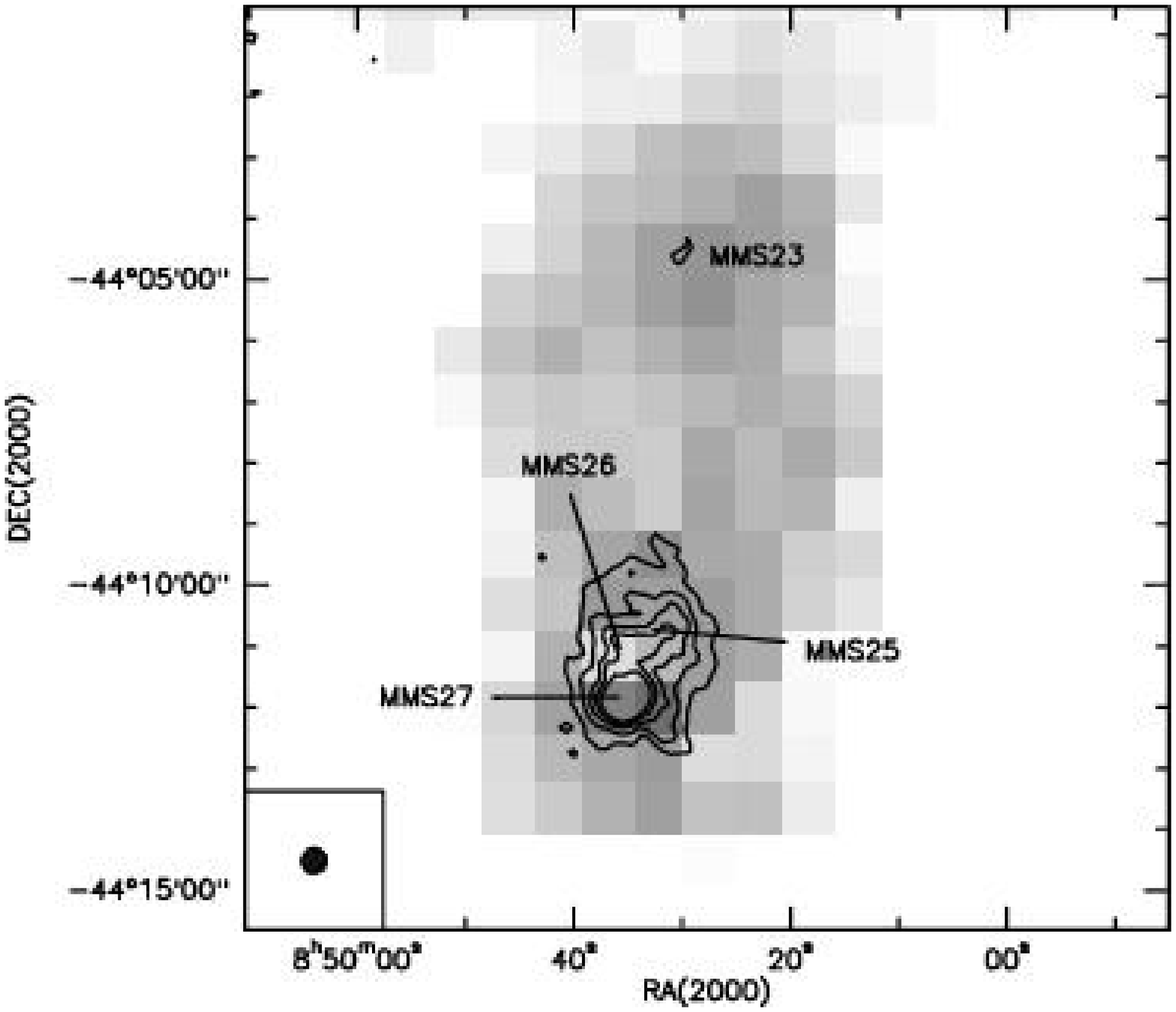}
      \caption{Contour map
      of the 1.2-mm continuum emission from the region including IRS21
      (area 4 in Fig.~\protect\ref{fig:mm_map}\protect).
      Contours are in steps 
      of 60 mJy/beam ($\sim 3\sigma$) from 60 to 300 mJy/beam.
      The cores are labelled and the SEST beam is
      drawn in the lower left box.
      Underlying, the $^{12}$CO(1--0) emission integrated
      from 0 to 20 km s$^{-1}$ (greyscale, from 30 to 150
      K km s$^{-1}$). 
         \label{fig:irs21}}
   \end{figure}
%
%_____________________________________________________________

\begin{figure}
   \centering
   \includegraphics[width=9cm]{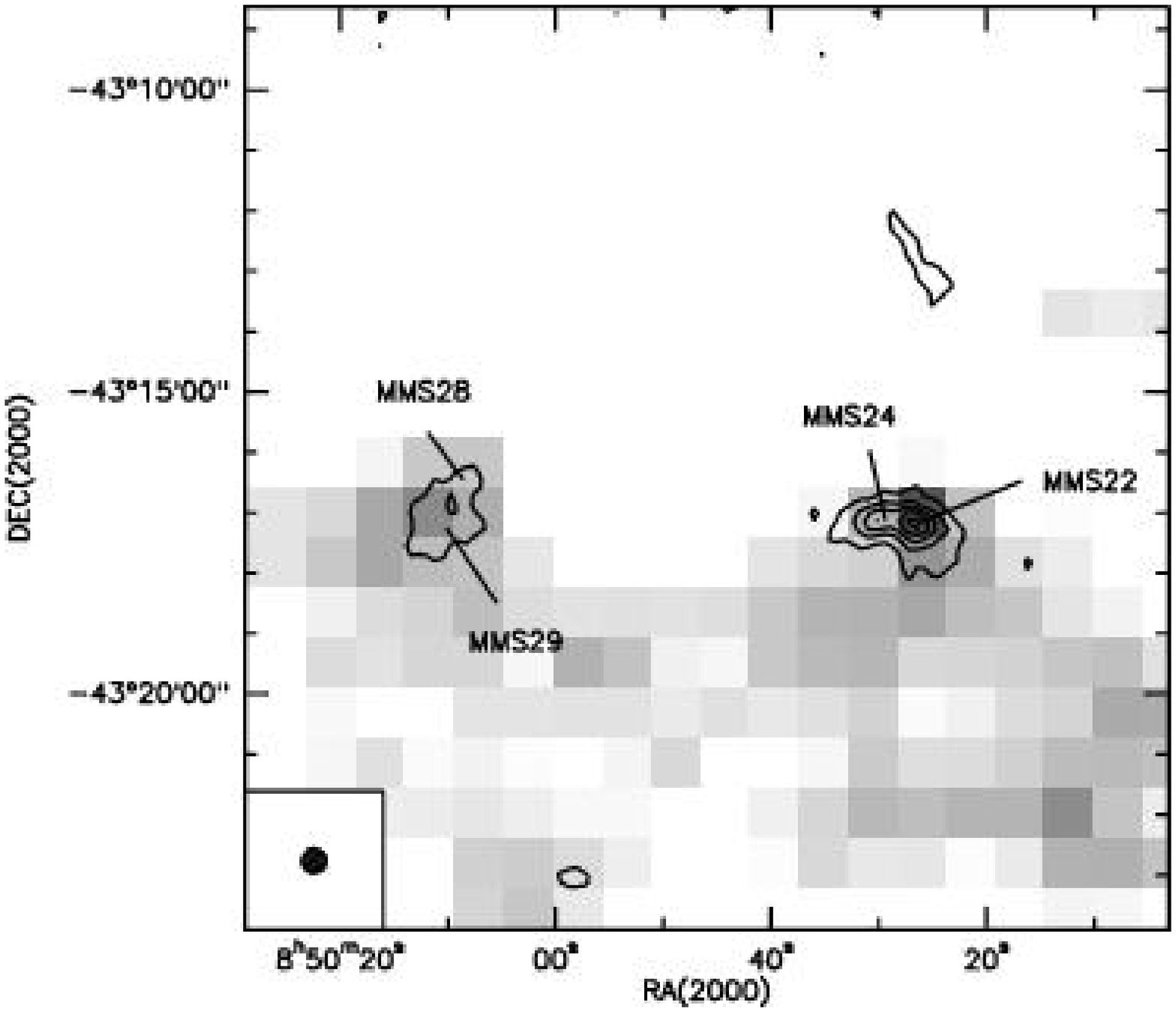}
      \caption{Contour map
      of the 1.2-mm continuum emission from the region including IRS20
      (area 5 in Fig.~\protect\ref{fig:mm_map}\protect).
      Contours are in steps
      of 120 mJy/beam ($\sim 6\sigma$) from 60 to 540 mJy/beam.
      The cores are labelled and the SEST beam is
      drawn in the lower left box.
      Underlying, the $^{12}$CO(1--0) emission integrated
      from 0 to 20 km s$^{-1}$ (greyscale, from 30 to 150
      K km s$^{-1}$). 
         \label{fig:irs20}}
   \end{figure}

\begin{figure}
   \centering
   \includegraphics[width=9cm]{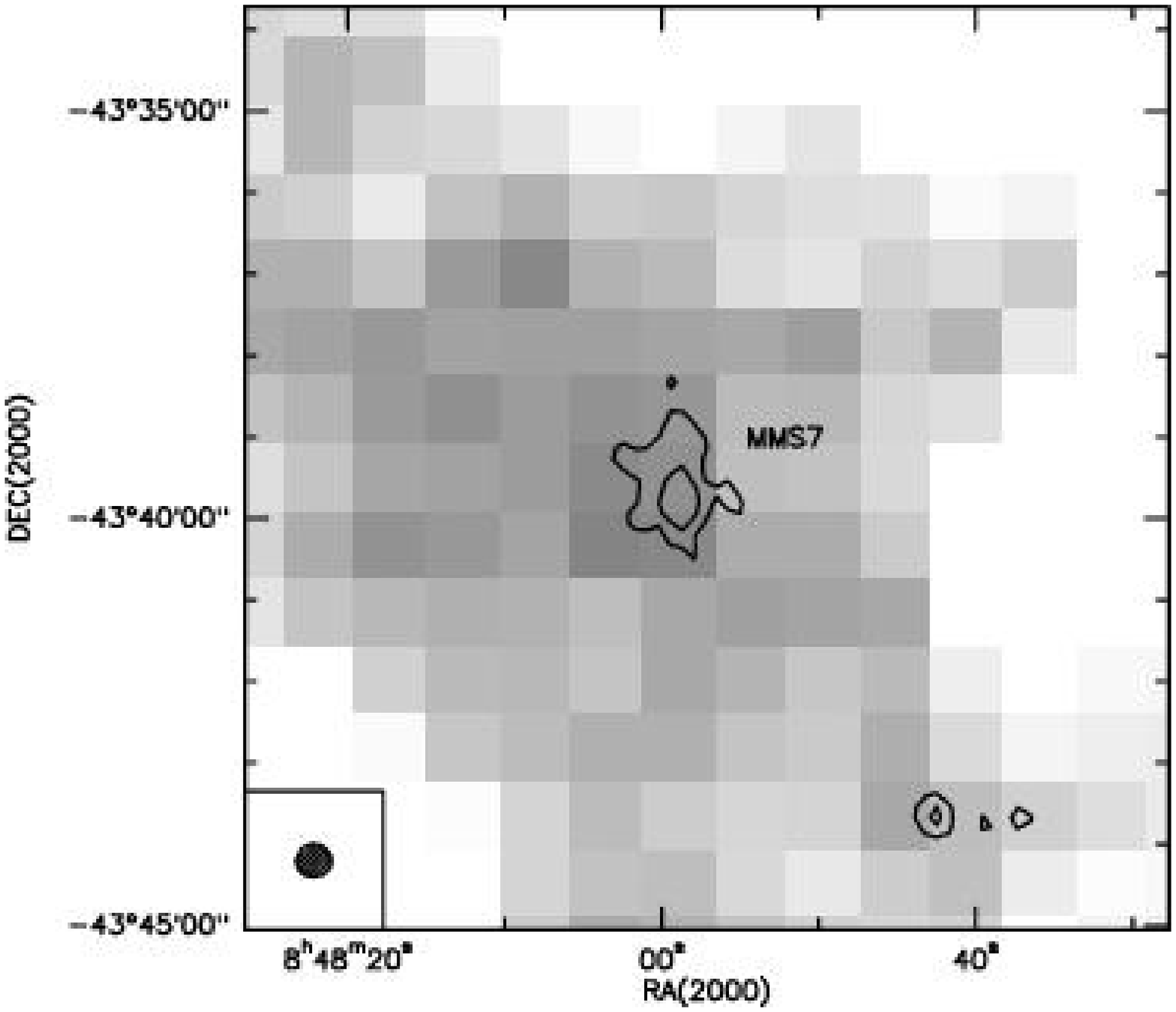}
      \caption{Contour map
      of the 1.2-mm continuum emission from the region at the
      centre of the mapped area (6 in Fig.~\protect\ref{fig:mm_map}\protect) 
      Contours are in steps 
      of 60 mJy/beam ($\sim 3\sigma$) from 60 to 120 mJy/beam.
      The cores are labelled and the SEST beam is
      drawn in the lower left box.
      Underlying, the $^{12}$CO(1--0) emission integrated
      from 0 to 20 km s$^{-1}$ (greyscale, from 30 to 150
      K km s$^{-1}$). 
         \label{fig:center}}
   \end{figure}

IRS21 (see Fig.~\ref{fig:irs21}) is the most massive in a chain of cores,
also aligned in the same direction as a surrounding filament of molecular
gas (i.\ e., north-south). 
IRS20 (see Fig.~\ref{fig:irs20}), as well, is part of a chain of
cores aligned with a surrounding filament of molecular
gas (i.\ e., east-west). Finally, a moderately-massive isolated core
lies towards the centre of the molecular cloud (see Fig.~\ref{fig:center}).

%_____________________________________________________________
%                FIGURA 6: IRS 20 
%-------------------------------------------------------------
%
%_____________________________________________________________

%_____________________________________________________________
%                FIGURA 7: CENTER 
%-------------------------------------------------------------
%
%_____________________________________________________________

\section{Discussion}
\label{discu}

\begin{figure*}
   \centering
   \includegraphics[width=10cm]{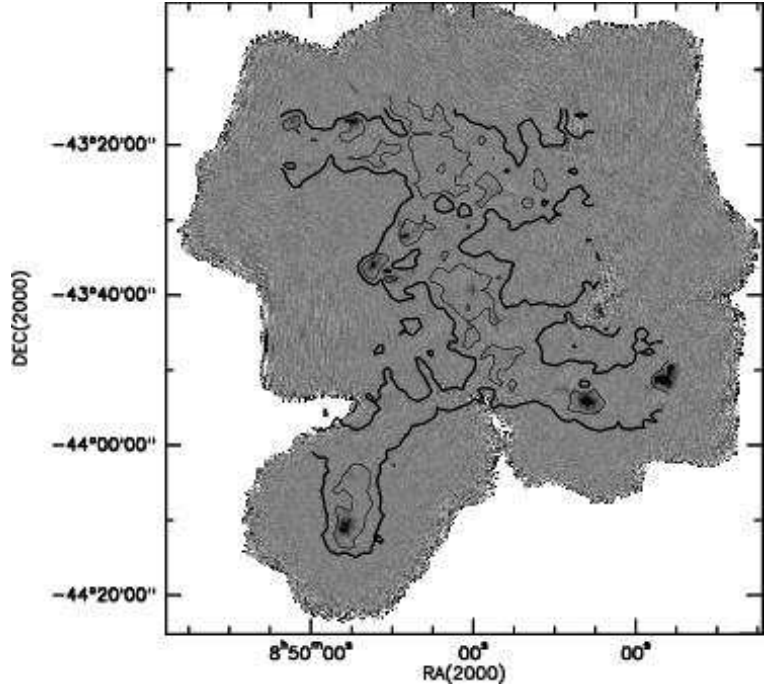}
      \caption{Map of CO(1--0) emission from the observed area
      within cloud D
      (contours), integrated from 0 to 20 km s$^{-1}$,
      obtained with a beam of $43\arcsec$. This is 
      overlaid with the SIMBA 1.2 mm continuum map (greyscale, beam of
      $24\arcsec$). The thick solid line marks the contour 
      at 25 K km s$^{-1}$, whereas the light solid line marks the
      contour at 60 K km s$^{-1}$.
         \label{fig:co_mm_map}
              }
   \end{figure*}

In Fig.~\ref{fig:co_mm_map}, we show the 1.2 mm continuum map overlaid with
contours of the map of CO(1--0) emission integrated from 0 to 20 km s$^{-1}$
(obtained with a beam twice as large as that at 1.2 mm). Clearly,
the molecular emission arises from filaments and the detected continuum
sources lie towards peaks of CO(1--0) integrated emission within those
filaments. This indicates that the mm cores all belong to the Vela D
cloud. In particular, the brightest of them are those associated
with the red IRAS sources and embedded clusters (IRS16, 17, 19, 20 and 21).
The kinematic distances attributed to some of these regions vary
in the range 0.6--1.8 kpc (see, e.g., Beltr\'{a}n et al.\ \cite{beltran},
Fa\'{u}ndez et al.\ \cite{faundez}); 
this is due to the structure evidenced by
the CO(1--0) spectra. Figure~\ref{fig:co_mm_map} shows that most filaments resemble 
arcs, and velocity-position cuts also give evidence for shell-like structures 
with differences in velocity 
of $\sim 5-10$ km$^{-1}$ s$^{-1}$. This suggests that the emission actually arises from the
same cloud, but that it is fragmented into a number of expanding shells. 
Hence our choice to attribute to all cores the same distance (700 pc). The issue
is discussed in detail in Elia et al.\ (\cite{elia}). 

%
%   FIG. 8: MAPPONE SOVRAPPOSTO A CO(1--0) 
%%%%%%%%%%%%%%%%%%%%%%%%%%%%%%%%%%%%%%%%%%%%
%%%%%%%%%%%%%%%%%%%%%%%%%%%%%%%%%%%%%%%%%%%%
%_____________________________________________________________

\begin{figure}
   \centering
   \includegraphics[width=9cm]{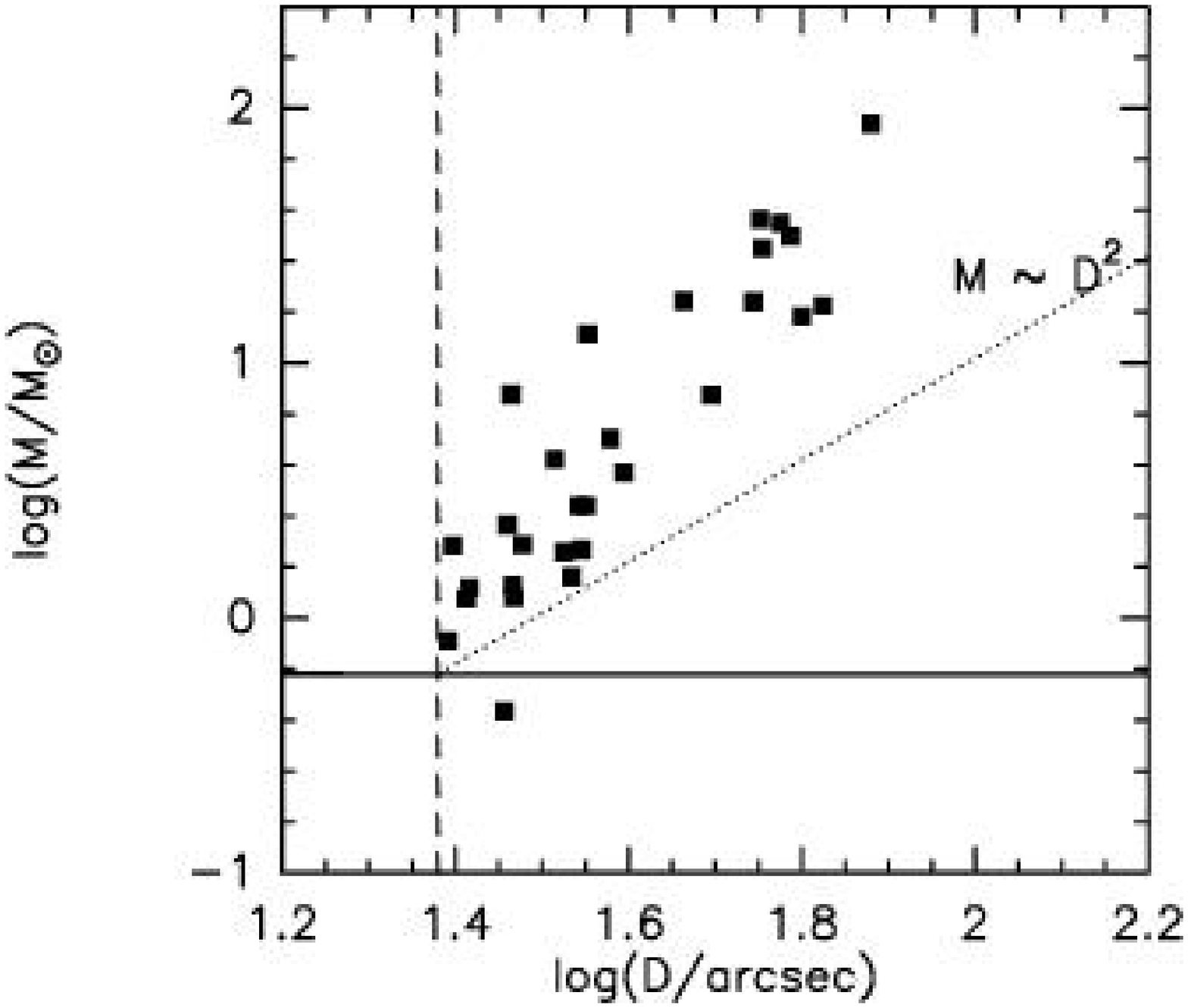}
      \caption{Cores' mass vs.\ size (uncorrected for beam).
      The vertical dashed line marks the SEST beam size, whereas the
      horizontal solid line shows the 3--$\sigma$ detection limit
      for point sources. 
      A dotted line has also been drawn following a $D^{2}$ relation,
      i.\ e., the sensitivity limit to extended sources (see text).
         \label{fig:logm-logd}}
   \end{figure}

The observed mass-versus-size relationship 
(uncorrected for beam) is plotted in Fig.\ref{fig:logm-logd}. 
The ``under-resolved''
detections reported in Table~\ref{table:umms} fall below the resolution limit given by the
beam size (the vertical line in figure).
%About half of the cores originally found by CLUMPFIND fall below the
%resolution limit given by the beam size (the vertical line in figure) 
%and in part might be false detections, although some of them are in the 
%neighborhood of far infrared point sources (correlations with FIR-NIR 
%sources will be presented in a forthcoming paper). The latter are
%listed in Table~\ref{table:cores}, indicated as mms\# (instead of
%MMS\#).  A higher sensitivity and resolution 
%is auspicable to resolve these emission regions and to
%detect possible multiple components of the cores (a plausible reason 
%for their non gaussian morphology).

%_____________________________________________________________
%                FIGURA 9: logD vs. logM 
%-------------------------------------------------------------
%
%_____________________________________________________________

\begin{figure}
   \centering
   \includegraphics[width=9cm]{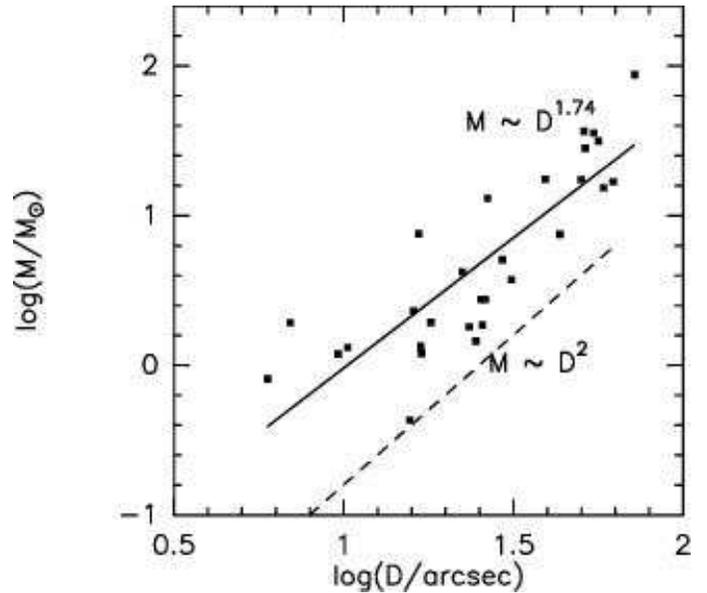}
      \caption{Cores' mass vs.\ deconvolved size. 
      The solid line is obtained by a fit ($M \sim D^{1.74}$),
      whereas the dashed line follows a $D^{2}$ relation,
      i.\ e., the sensitivity limit to extended sources (see text).
         \label{fig:logm-logd:dec}}
   \end{figure}

When assessing the sensitivity of a survey in the mm continuum, one has to
take into account the extent of the emitting area. Since the sensitivity
limit is a flux density per solid angle, for a given total flux density
an extended source is less likely to be detected than a point source.
In fact, assuming constant dust opacities and temperatures, the source
mass is proportional to the {\em total} (i. e., integrated over the
solid angle) flux density. 
Then, the {\em mass} sensitivity limit is given by the
flux sensitivity per solid angle multiplied by the source area and so increases
with increasing areas. 
This is illustrated in Fig.~\ref{fig:logm-logd}: all sources lie above a
line following a $M \sim D^{2}$ relation (where $D$ is the core size), which 
represents the actual sensitivity limit. Clearly, the data points outline
a steeper dependence of mass vs. size; a linear fit indicates a $M \sim D^{3.7}$
relation.  This appears to be intrinsic to
the core population, indicating that the sample is
likely to be complete at least down to $\sim 1 - 1.3$ $M_{\odot}$. 
A more significant relation is obtained by fitting deconvolved sizes
vs. mass 
(see Fig.~\ref{fig:logm-logd:dec}), 
yielding $M \sim D^{1.7}$. This is flatter than 
found for pre-star forming regions in the range $0.01-10$ pc by,
e.g., Heithausen et al.\ (\cite{heit}) using CO emission
($M \sim D^{2.31}$), but steeper than found for pre-stellar
cores by Motte et al.\ (\cite{motte01}) using submm emission
($M \sim D^{1.1}$). Instead, Reid \& Wilson (2005) find
$M \sim D^{x}$ with $x \sim 1.5-2.1$ in the high-mass star forming region
NGC7538 from submm observations, more in agreement with our result.
However, note that 9 out of the 29 sources in our robust sample are 
smaller than the beam FWHM, i. e., their beam-convolved size is
$< 1.5$ FWHM. Then, the core size may become highly uncertain 
at the lowest end of the distribution. 
%This, in turn, might affect the
%derived mass-vs-size relation, although probably not in a significant
%way. 

%_____________________________________________________________
%                FIGURA 9 bis: logD vs. logM DECONVOLUTO 
%-------------------------------------------------------------
%
%_____________________________________________________________
%

\begin{figure}
   \centering
   \includegraphics[width=9cm]{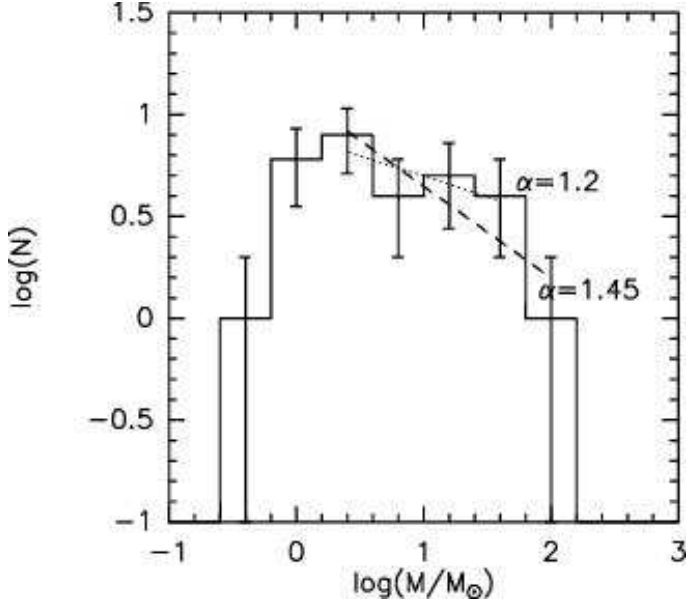}
      \caption{Mass spectrum of the mm cores. 
      The errorbars indicate the r.m.s. assuming a
      Poisson statistic. A dashed
      line shows a ${\rm d}N/{\rm d}M \sim M^{-1.45}$
      relation, 
      a dotted line shows a ${\rm d}N/{\rm d}M \sim M^{-1.2}$
      relation.  The estimated completeness limit is
      $\sim 1-1.3$ $M_{\sun}$ (see text).
         \label{fig:mass}}
   \end{figure}

The core mass spectrum is shown in Fig.~\ref{fig:mass}. The errorbars
are the r.m.s. estimated by assuming a Poisson statistic. The distribution
appears quite flat and drops abruptly at $\sim 10^2$ $M_{\sun}$. A power-law 
fit to the points {\em above} the completeness limit indicates that
the mass spectrum follows a relation ${\rm d}N/{\rm d}M \sim M^{-\alpha}$,
with $\alpha = 1.45 \pm 0.2$ ($\chi^{2} = 1.10$). This is shallower than a Salpeter-like
Initial Mass Function  ($\alpha \sim 2.35$), being more similar
to the mass spectrum of molecular cores in pre-star forming regions
(see, e.g., Heithausen et al.\ \cite{heit}). 
However, Fig.~\ref{fig:mass} suggests that fitting the core mass function with
two power-law segments 
would be more adequate. In this case, we obtain $\alpha = 1.20 \pm 0.34$ 
($\chi^{2} = 0.42$) up
to $M \sim 40$ $M_{\sun}$, with a significantly
larger $\alpha$ for $M > 40$ $M_{\sun}$. 
As for (sub-)mm surveys, the derived spectral index roughly
agrees with that determined by Kerton et al.\
(\cite{kerton}) for KR140 between $\sim 0.7$--100 $M_{\sun}$
($\alpha = 1.49$), and by Mookerjea et al.\ (\cite{mookerjea}) 
between 37--16000 $M_{\sun}$ in R106 ($\alpha \sim 1.5-1.7$). 
Reid \& Wilson (\cite{reid}) also find a flat distribution ($\alpha = 0.9
\pm 0.1$) between $\sim 1-100$ $M_{\sun}$ in NGC7538, then steepening
to $\alpha = 2.0-2.6$. 

%_____________________________________________________________
%                FIGURA 10: logN vs. logM 
%-------------------------------------------------------------
%
%_____________________________________________________________

The slope of the mass spectrum and the mass-size
relationship are not sensitive to the adopted dust opacity,
temperature and distance as long as {\em all} cores are characterized by the same
{\em mean} opacity and temperature and lie at the same distance. As discussed before,
both differences in distance and contributions from free-free and synchrotron
radiation to the dust thermal emission
do appear to be minor issues. Large errors may only be caused by having
detected thermal dust emission coming from environments whose temperatures
and/or dust opacities differ within the sample. In fact, the identified 
cores are likely to be a mixture of protostellar envelopes (those coinciding with 
the red IRAS
sources), pre-protostellar cores/Class 0 sources and transient structures. 
A forthcoming paper (De Luca et al.\ \cite{delu07}) will
discuss the coincidence between mm cores and NIR/MIR/FIR sources and the nature
of the single cores. Here, just to test what would be the effect of 
gathering together sources with different temperatures and opacities, 
we have subdivided the robust sample into a subset of objects (10 out of 29) 
that may be associated with IRAS or MSX sources and a subset of objects unassociated
with IRAS or MSX sources (the remaining 19). To the first group we have
attributed a temperature of 30 K and a dust opacity $k_{1.3} = 1$ cm$^{2}$ g$^{-1}$,
whereas to the second group we have attributed a temperature of 15 K and a dust 
opacity $k_{1.3} = 0.5$ cm$^{2}$ g$^{-1}$ (as usually found in the literature for
pre-protostellar cores; e. g., Motte et al.\ \cite{motte}, \cite{motte01},
Nutter et al.\ \cite{nutter}). 
By only adopting two different temperatures (i. e., using a same dust
opacity $k_{1.3} = 0.5$ cm$^{2}$ g$^{-1}$) we obtain a similar
mass spectrum ($\alpha = 1.44$) but a less steep mass-size relation
($M \sim D^{1.38}$), whereas two different temperatures and two different
opacities yield both a steeper mass spectrum ($\alpha = 1.89$) and
a shallower mass-size relation ($M \sim D^{1.12}$).

However, the fraction of cores where dense warm dust dominates
may be lower than envisaged above.
See, e.g., Fig.~\ref{fig:irs16}: the cores MMS1, 2 and 3 are relatively far
from the IRAS uncertainty ellipse, at the border of the HII region. 
Since the 1.2 mm emission from the area of
the optical/radio HII region appears higher than extrapolated from the radio flux 
density (see previous section) assuming optically thin free-free emission, 
possibly there exists also warm dust mixed with
the ionized gas, but this lies outside the cores. Consider the case
of IRS17, as well: as discussed in Giannini et al.\ (2005), see also
Fig.~\ref{fig:irs17}, the core found towards
the IRAS source (MMS4) encompasses both the IRAS uncertainty ellipse and a less
evolved protostar (at the origin of a prominent NIR jet). Then,
although it was included in the subset of sources with higher temperature and dust opacity,
it is actually possible that a major contribution comes from cold dust surrounding 
the protostar (i. e., with lower temperature and dust opacity). 
So, a large fraction
of the 1.2 mm emission from IRS16 and IRS17 may be actually due to colder 
dust and this
might be true for the whole subset of cores to which a higher temperature and 
dust opacity have been attributed in running this test.
Nevertheless, the mass spectra and size-mass relations as derived in the two cases
of uniform and
different mixtures of environments probably bracket the actual ones. 
 
A core mass spectrum with $\alpha \sim 1.4 - 1.9$ 
(or $\alpha \sim 1.2$ up to $\sim 40$ $M_{\sun}$) is consistent 
with the clump mass spectra in the same region
obtained from CO(1--0) and $^{13}$CO(2--1) by Elia et al.\ (\cite{elia}),
exhibiting spectral indices $\alpha \sim 1.3 - 2.0$. This may suggest
a common origin for clumps and cores; however, Elia et al.\ (\cite{elia}) discuss
a number of problems in the derivation of the mass spectra from line observations
and also find steeper mass-size relationships.

%
% CORREZIONE 1
%
%A ``flat'' mass-size relation ($M \sim D$) 
A ``flat'' mass-size relation 
%often taken as an indication that the cores may represent 
%a sample of self-gravitating Bonnor-Ebert condensations
has been shown to be consistent with
a sample of pressure-confined Bonnor-Ebert condensations exhibiting 
uniform temperature and a range of external pressures (e.g., Motte
et al.\ \cite{motte01}, Johnstone et al.\ \cite{john}). 
Whether the observed mass-size relation is due to all cores being 
close to the critical Bonnor-Ebert sphere relation, in which case $M
\propto D$ for a range of pressures, or whether it is formed by
families of Bonnor-Ebert spheres with varying importance of gravity,
in which case $M \propto D^{3}$ for any fixed pressure, is uncertain 
at the sensitivity limits of present surveys.
%
% FINE CORREZIONE 1
%
%
% CORREZIONE 2
%
%On the contrary, a steeper 
A mass-size relation $M \sim D^{2}$ would be 
%
% FINE CORREZIONE 2
%
more consistent with turbulent molecular 
cloud fragmentation producing a fractal distribution of core masses and sizes
(e.g., Elmegreen \& Falgarone \cite{elme}). Actually, a sample of singular isothermal
spheres (e.g., Shu et al.\ \cite{shu}) with constant sound speed and different masses 
would exhibit a $M \sim D$ relation, as well. 
However, this would imply that the clumps are all gravity bound, otherwise a much larger
variation of the external pressure would be required to confine them with respect
to Bonnor-Ebert condensations.
%
% INIZIO CORREZIONE 7
%
%it is easy to check that at later evolutionary stages, 
%this sample would tend to produce a $M \sim D^{1.5}$ relation (i.\ e., for
%a $r^{-3/2}$ density distribution within the cores). 
%{\em
%A sample of singular isothermal sphere undergoing gravitational collapse
%(then approaching a $r^{-3/2}$ density distribution)
%would still tend to produce a $M \sim D^{1}$ relation.
%} 
%
% FINE CORREZIONE 7
%

The mass-size relation we
derive ($M \sim D^{1.7}$) is shallower than that predicted for molecular cloud 
turbulent fragmentation,
%
% CORREZIONE 3
%
%but steeper than that expected for Bonnor-Ebert condensations or singular isothermal
%spheres. 
with the exponent in between those expected for the different families of
Bonnor-Ebert spheres.
%
% CORREZIONE 3
%
However, we showed that 
%
% CORREZIONE 4
%
%this value 
the actual exponent 
%
% FINE CORREZIONE 4
%
is sensitive to the fraction
of warm dense gas within the sample, decreasing with increasing this fraction. 
When crudely accounting for this effect, the exponent lies between
1.5 and 1, ruling out turbulent fragmentation and 
suggesting that our sample includes gas condensations in different evolutionary
stages. High resolution measurements of the cores' column density profiles would
be critical to clarify the issue. 

The core mass spectrum between $\sim 1$ and $\sim 100$ $M_{\sun}$ is shallower 
than the stellar IMF in the same mass range, even in the worst case of a mixture 
of cores of different densities and temperature.  Massi et al.\ (\cite{massi06}) find a
standard stellar IMF for a sample of 6 young star clusters in the D cloud (5 of them associated
with observed mm cores), hence the core mass spectrum is more reminiscent of the
mass spectrum of molecular clumps, as already noted. This means that the larger cores 
underwent 
fragmentation, but this has not been observed at mm wavelength because of low resolution. 
The core mass spectrum should then be compared with the {\em cluster} mass spectrum 
%
% INIZIO CORREZIONE 8
%
(i. e., the mass distribution of stellar clusters)
%
% FINE CORREZIONE 8
%
rather than the stellar IMF.
Only at the lower mass end, the core mass spectrum may be dominated by the 
precursors of lower mass {\em isolated} stars and is likely to approximate an IMF. 
Remarkably, Lorenzetti et al.\ (\cite{lorenzetti}),
based on FIR fluxes from the IRAS Point Source Catalogue, derived a mass distribution
with an index $\sim 1.5$ for the brightest
protostars in the Class I stage associated with the Vela Molecular Ridge.
As found by Massi et al.\ (\cite{massi99}),
although the most luminous IRAS sources coincide with embedded young clusters,
it is the most massive star of each cluster that appears to dominate the FIR emission.
If so, the similarity between core mass spectrum and Class I source mass spectrum
suggests that the mass of the most massive stars originated by a core is roughly
proportional to the core's mass itself, at least at the high mass end.
So, the core mass spectrum would reflect the mass spectrum of the most massive
stars of every stellar group and, probably, the cluster mass spectrum, whereas
the stellar IMF would be dominated by the fragmentation undergone by the single
cores. Note that if the number of clusters' members and the mass of the most massive star
in a cluster both depend on the mass of the parental core, then there must also exist 
an observable relationship between number of clusters' members and mass of the most 
massive star
in a cluster. This would confirm that small stellar clusters cannot produce
massive stars, explaining the possible lack of massive stars in the region as
suggested by the results of the NIR observations by Massi et al.\ (2006). 
 
%
% CORREZIONE 9
%
It is straightforward to derive a gas density $n({\rm H_{2}}) \sim 10^{5}$ cm$^{-3}$ 
for MMS4, the most
massive mm core. A mass-size relation $M \sim D^{1.7}$ implies a 
density-size relation $n({\rm H_{2}}) \sim D^{-1.3}$, so it is easy to check that
the density of mm cores is $> 10^{5}$ cm$^{-3}$. 
%
% FINE CORREZIONE 9
%
We can estimate the fraction of gas mass in dense 
(i. e., $n({\rm H_{2}}) > 10^{5}$ cm$^{-3}$) cores over the total gas mass
by summing up all masses
listed in Table~\ref{table:cores}; as a whole, we obtain $\sim 355$ $M_{\sun}$
of gas from 1.2 mm emission. Using our CO observations,
the cloud mass ranges from $1.5 \times 10^{4}$ $M_{\sun}$
(based on the $^{12}$CO integrated emission) to $3.7 \times 10^{3}$ $M_{\sun}$
(based on the LTE method and roughly correcting for the area not observed
in $^{13}$CO), hence $\sim 2.4 - 10$ \% of the gas lies in dense cores.
The lower end is comparable, e.g., to the mass fraction of starless and
prestellar cores of the most active star forming region in L1688 
(the $\rho$ Oph.\ main cloud)
and L1689, but is larger than the ``total'' mass fraction of starless
and prestellar cores in the two clouds (Nutter et al.\ \cite{nutter}). 
Johnstone et al.\ (2004) find that the submillimetre objects represent only
a 2.5 \% of the mass of the Ophiuchus cloud. Kirk et al.\ (2006) surveyed the
Perseus molecular cloud and derived a gas mass contained in submillimetre objects 
amounting to a 1 \% (or even less) of the total cloud mass, 
that seems consistent with the results of Hatchell et al.\ (2005).
Hence, only few percent of the mass of molecular clouds appears to be in the form
of dense cores and it is difficult to assess whether this mass fraction may
be related with global properties of the parental clouds. 

By comparing the total mass of dense gas associated with each embedded star
cluster and the total stellar mass estimated by Massi et al.\ (\cite{massi06}),
we find that in IRS16, IRS17 and IRS 21 the mass is almost equally distributed
between dense gas and stars, whereas in IRS19 and IRS21 the stellar mass
is larger than that of dense gas. Hence, for IRS16, IRS17, IRS19 and IRS21,
the star formation efficiency computed by using 1.2 mm continuum emission ranges
between 50 and 66 \%, whereas that computed by using line emission 
(according to Massi et al.\ \cite{massi06}) appears lower, 10--26 \%. 
Obviously, the latter takes account of less dense gas not detected in the
continuum. In the case of IRS20, in contrast, the star formation efficiencies are similar
(80 and 77 \%). Towards all of the 5 clusters, 
there are in principle still sufficiently massive cores
to produce high-mass stars, provided a high efficiency is possible in converting 
dense gas into a star. Outside the clusters, massive cores have not been found and 
only intermediate- and low-mass stars can form.

An extinction threshold for star formation has been suggested by several authors.
E.g., Onishi et al.\ (\cite{onishi}) and Johnstone et al.\ (\cite{john04}) 
find $A_{V} \sim 7 - 9$ mag in Taurus and Ophiuchus. As shown in 
Fig.~\ref{fig:co_mm_map}, there is quite a good correspondence of CO(1--0)
integrated intensity enhancements and mm cores, and CO(1--0) is a good tracer
of diffuse gas. Then, an $A_{V}$ can be estimated by converting CO(1--0)
integrated emission into $N({\rm H}_{2})$, using the empirical
relation of Elia et al.\ (\cite{elia}). 
No mm peaks are detected at $A_{V} < 12$ mag, which compares
well with the values found in Taurus and Ophiuchus, given the uncertainty
in the conversion of CO(1--0) integrated emission into H$_{2}$ column density.
However, we showed that our survey is probably not complete below 1 $M_{\sun}$
and so low-mass cores may have escaped detection.

%Confronto molto rapido con la mappa preliminare del CO(1--0) basato su una versione
%migliorata di Fig.~\ref{fig:mm_map}. Poi potremmo introdurre delle figure
%tipo Fig.~\ref{fig:irs16} e Fig.~\ref{fig:irs19} in cui si confrontano
%immagini IR dei clusters e dettagli dell'emissione a 1.2mm. In particolare,
%Fig.~\ref{fig:irs16} mostra che i clumps sono ai bordi della regione HII e,
%quindi, probabilmente NON sono free-free ma polvere. Fig.~\ref{fig:irs19}
%mostra bene una struttura filamentosa. Per IRS 17 possiamo riferirci all'articolo
%o possiamo riproporre una versione aggiornata della figura con cluster e clump.
%Discussione sui filamenti (tipo Hatchell et al. 2005 ``Star Formation in Perseus'').
%Mi piacerebbe anche discutere brevemente la sostanziale uguaglianza tra
%l'indice spettrale derivato da Lorenzetti et al. (1993) per le Classi I in Vela
%e quello che troviamo sulla base dell'emissione a 1.2mm.
%FRAZIONE DI MASSA all'interno dei clumps!!

To assess the nature of the observed filaments, we attempted the analysis
made by Hatchell et al.\ (\cite{hatchell}), which is based on the
work by Fiege \& Pudritz (\cite{fiege}). We considered the
filaments composed of mm cores MMS5 and MMS6 (east of IRS17),
MMS8 and MMS9, MMS10 and MMS11, MMS17 and MMS18, MMS19, MMS20 and MMS21
(close to IRS19). All have a small mass per unit length, ranging
from $\sim 5$ to $\sim 40$ $M_{\sun}$ pc$^{-1}$. The virial mass
entering Eq.~11 of Fiege \& Pudritz (\cite{fiege}) was derived
through their Eq.~12 by using the FWHM of $^{13}$CO(2--1) lines observed
towards the mm cores. The FWHM's span the interval 1.2--2.8 km s$^{-1}$,
increasing towards cores associated with IRAS sources (with the
exception of MMS5 and MMS6, not associated with IRAS sources but
exhibiting a large FWHM, i. e. $2.3$ km s$^{-1}$). The velocity dispersion ($\sigma$)
entering the virial mass has been determined using Eq.~20 and Eq.~21
of Fiege \& Pudritz (\cite{fiege}), assuming a gas temperature of 30 K. 
We obtained virial masses $m_{\rm vir}$
in the range $\sim180$ to $\sim 500$ $M_{\sun}$ pc$^{-1}$, in all cases at least
one order of magnitude more than the observed masses per unit length.
Since the filaments lie well within the larger filaments of molecular
gas, we derived the external pressure $P_{\rm S}$ on the cores using
Eq.~2.12 of McKee (\cite{mckee}). Given the average column density
is $\sim 10^{22}$ H$_{2}$ cm$^{-2}$ towards the 
gas encompassing the denser filaments,
the pressure over the cores' surface is $\sim 5 \times 10^{5}$ K cm$^{-3}$.
Assuming a density of $10^{5}$ cm$^{-3}$, the cores' inner pressure
is given by $P = \sigma^{2} \rho$, i.\ e., in the range $1-5 \times 10^{7}$
K cm$^{-3}$. Following the analysis of Fiege \& Pudritz (\cite{fiege}),
in particular with reference to their Eq.~11,
the filaments cannot be confined by the external pressure and
a toroidal magnetic field is needed to confine them. 
This is consistent with what found by Elia et al.\ (\cite{elia}) for molecular
clumps, i. e.\ the clumps' virial masses are always about an order
of magnitude larger than the observed ones. That depends in turn on the
derived velocity dispersion.  However, the velocity
dispersion as determined from $^{13}$CO(2--1) is very likely not
the correct one within the filaments. $^{13}$CO(2--1) traces 
the lower density gas outside the dense cores and is probably much more sensitive to 
the star formation activity that is ongoing in the region.
Hence, mm observations of high-density tracers and an assessment of the
evolutionary status of the filaments through a comparison with near- and
mid-IR data is necessary to analyse the stability of the filaments. 
In conclusion, the present data do not allow us to ascertain whether
the fainter filaments are pressure- or magnetically confined,
are collapsing structures or are dispersing.

\section{Conclusions}
\label{conclu}

The discussion above suggests the following scenario for star
formation activity in the observed molecular cloud. The arcs
and double-peaked spectra evidenced by the CO(1--0) data suggest
that the current gas distribution has its origin in 
intersecting expanding shells. The shells fragmented producing large filaments with
embedded dense clumps. Within the clumps themselves, filaments of dense gas 
(less than 10 \% of the total gas mass)
fragmented giving rise to the dense cores observed in the mm
continuum. The most massive cores are associated with embedded
young stellar clusters, indicating further fragmentation.
The efficiency with which dense gas is converted into stellar
mass appears high, $>50$ \%. Furthermore, the dynamical picture
would confirm the widespread view according to which cluster
formation is triggered by external
compression of the parental gas (for a brief review, see
Tan \cite{tan}).
Massi et al.\ (\cite{massi06}) infer ages $> 10^{6}$
yr for the embedded clusters, which is consistent with the shells' 
dynamical age (see Elia et al.\ \cite{elia}).
Hence, we are probably sketching star formation occurring
in the last few Myr. The core mass spectrum appears flatter than 
a standard IMF in the range $\sim 1-100$ $M_{\sun}$, but
agrees with the typical mass spectra of molecular clumps. 
Flat mass spectra have also been obtained from CO observations
of the same region.  It also
agrees with the mass distribution of relatively luminous Class I sources
in the region, indicating a continuity between molecular clumps, dense cores
and the most massive stars in each cluster. The core
mass spectrum should also be compared with the cluster mass spectrum
in the region. Since, as found by Massi et al.\ (\cite{massi06}),
the stellar IMF in the clusters appears similar to a standard IMF, it is likely
that the ``global'' IMF results from the bulk of clusters' members,
accounting for the apparent discrepancy between the cores mass spectrum and a
standard IMF. Massi et al.\ (\cite{massi06}) find also a lack 
of high-mass stars in relation to what would be expected from a standard IMF.
As pointed out by Lorenzetti et al.\ (\cite{lorenzetti}),
no massive star formation is currently occurring in the region. 
This would be the case if both the mass of the most massive member
in a cluster and the number of clusters' members were roughly proportional
to the mass of the parental cores, as suggested by the similarity
between mass spectra.  However,
a few cores are massive enough to produce high-mass stars.
To assess this possibility, further high-resolution multi-wavelength observations
are needed to delineate the actual evolutionary stage of the cores. 

In summary, the main quantitative results are:
 \begin{enumerate}
      \item the core mass spectrum follows a relation ${\rm d}N/{\rm d}M 
        \sim M^{-\alpha}$ between $\sim 1 -100$ $M_{\sun}$,
        with $\alpha \sim 1.45 - 1.9$ depending on the temperature and opacity 
       distribution within the sample; if a two-segments power-law is fitted, the 
       power-law index is
     $\alpha = 1.2$ up to $\sim 40$ $M_{\sun}$ and appears significantly larger beyond;
      \item a mass-size relation $M \sim D^{x}$ has been derived for the
         cores, with $x \sim 1.1 - 1.7$, again depending on the
         temperature and opacity distribution within the sample;
      \item no cores have been detected in areas with $A_{V} < 12$ mag,
         a possible confirmation of the existence of an extinction threshold
          for star formation;  
      \item the filaments of dense gas towards some of the cores cannot
        be confined by external pressure, but this result critically 
        depends on having used the velocity dispersion as inferred 
        from $^{13}$CO(2--1) observations.
 \end{enumerate}

%\begin{acknowledgements}
%
%\end{acknowledgements}

\end{document}